# High CO$_2$-tolerance oxygen permeation dual-phase membranes Ce$_{0.9}$Pr$_{0.1}$O$_{2-\delta}$-Pr$_{0.6}$Sr$_{0.4}$Fe$_{0.8}$Al$_{0.2}$O$_{3-\delta}$

Lei Shi[a], Shu Wang[a], Tianni Lu[b], Yuan He[a], Dong Yan[a], Qi Lan[a], Zhiang Xie[a], Haoqi Wang[a], Man-Rong Li[c], Juergen Caro[d], Huixia Luo[a]*

[a]*School of Material Science and Engineering and Key Lab Polymer Composite & Functional Materials, Sun Yat-Sen University, No. 135, Xingang Xi Road, Guangzhou, 510275, P. R. China*

[b]*School of Materials Sciences and Engineering, Shenyang Aerospace Unversity, Shenyang, 110136, P. R. China*

[c]*School of Chemistry, Sun Yat-Sen University, No. 135, Xingang Xi Road, Guangzhou, 510275, China*

[d]*Institute of Physical Chemistry and Electrochemistry, Leibniz University of Hannover, Callinstr. 3A, D-30167 Hannover, Germany*

*Corresponding author/authors complete details (Telephone; E-mail:) (+0086)-2039386124

<u>luohx7@mail.sysu.edu.cn</u>

**ABSTRACT**

High stability and oxygen permeability are two prominent requirements for the oxygen transport membrane candidates used as industrialization. Herein, we report several composite membranes based on $x$wt.%$Ce_{0.9}Pr_{0.1}O_{2-\delta}$ (CPO)-(100-$x$)wt.%$Pr_{0.6}Sr_{0.4}Fe_{0.8}Al_{0.2}O_{3-\delta}$ (PSFAO) ($x$ = 50, 60 and 75) prepared via a modified Pechini method. Oxygen permeability test reveals that the 60CPO-40PSFAO composition exhibits the highest oxygen permeability. The oxygen permeation flux through the optimal uncoated 0.33 mm-thickness 60CPO-40PSFAO composite can reach 1.03 mL cm$^{-2}$ min$^{-1}$ (over the general requirement value of 1 mL cm$^{-2}$ min$^{-1}$) in air/He atmosphere at 1000 °C. *In situ* XRD performance confirms the optimal 60CPO-40PSFAO sample shows excellent stability in $CO_2$-containing atmospheres. The 60CPO-40PSFAO membrane still exhibits simultaneously excellent oxygen permeability and phase stability after operating for over 100 h at air/$CO_2$ condition at 1000 °C, which further indicates that the 60CPO-40PSFAO composite is likely to be used for oxygen supply in $CO_2$ capture.

# 1. Introduction

Interest in ceramic oxygen transport membranes (OTMs) technology is growing exponentially across various scientific and engineering disciplines because of the widespread applications in the field of solid oxide fuel cells (SOFCs) [1, 2], oxygen separation [3-5], hydrocarbons conversion [6-8], oxy-fuel combustion for $CO_2$ capture and so forth. During the oxy-fuel process, high purity oxygen production has been put forward to use OTMs technology instead of the conventional cryogenic technique, which can benefit from low costs and energy consumption and high efficiency. However, the candidates of OTMs selected for the oxy-fuel process should possess good oxygen permeability and excellent thermochemical stability under corrosive gases (such as $CO_2$, $SO_2$, $H_2S$ etc.) conditions [9-14].

Among the various OTMs, single phase perovskite OTMs (e.g. $Ba_{1-x}Sr_xCo_{1-y}Fe_yO_{3-\delta}$ [11], $SrCo_{1-y}Fe_yO_{3-\delta}$ [12], $BaBi_xCo_yFe_{1-x-y}O_{3-\delta}$ [14]) with good oxygen permeability have been of high interest. However, their industrial practical application was limited since they suffer from low mechanical strength and poor stability in $CO_2$- or $SO_2$-containing atmospheres [15, 16]. In order to overcome the limitations of single perovskite materials, researchers have shifted to explore dual-phase OTMs, which are usually composed of an electron conductor (EC) and oxygen ion conductor (OIC). By far, driven by the demands for oxy-fuel concept, already many $CO_2$-stable dual phase membrane materials have been developed [17-27], such as $Ce_{0.8}Sm_{0.2}O_{1.9}$-$SmMn_{0.5}Co_{0.5}O_{3-\delta}$ (CSO-SMCO) [17], $Ce_{0.8}Sm_{0.2}O_{2-\delta}$-$La_{0.9}Sr_{0.1}FeO_{3-\delta}$ (CSO-LSFO) [18], and $Ce_{0.9}Nd_{0.1}O_{2-\delta}$-$Nd_{0.6}Sr_{0.4}FeO_{3-\delta}$ (CNO-NSFO) [19]. Although their $CO_2$ stabilities have been enhanced, most of their oxygen permeabilities are too low for practical applications.

In order to enhance oxygen permeability and/or improve material thermochemical stability of OTMs, the large majority of studies focus on tuning the composition by chemical doping [20-29]. Aluminum is one of the earth-abundant elements belonging to IIIA group with a fixed valence state. Thus, it has been classically used in the single perovskites-based OTMs due to its low cost and a more stable redox behavior of the material [30-40]. Unfortunately, most reports confirmed that the oxygen permeation rate decreased after doping aluminum in the single lanthanum-based or strontium-based perovskite oxides except for the $(Ba_{0.5}Sr_{0.5})(Fe_{1-x}Al_x)O_{3-\delta}$ and $Ba_{0.5}Sr_{0.5}Co_{0.8}Fe_{0.1}Al_{0.1}O_{3-\delta}$ system [30-41]. However, the oxygen permeability and the thermochemical stability under reducing atmospheres or corrosive (such as $CO_2$) atmosphere have both been enhanced with the substitution of Al at the *B*-site of OIC in the dual-phase OTM case [42-44]. More recently, J. Caro group have developed a novel $CO_2$-stable 60wt.%$Ce_{0.9}Pr_{0.1}O_{2-\delta}$-40wt.%$Pr_{0.6}Sr_{0.4}FeO_{3-\delta}$ (denoted as 60CPO-40PSFO) composite material [45]. Inspired by these previous studies [46-48], we propose to design and prepare a series of $Ce_{0.9}Pr_{0.1}O_{2-\delta}$ (CPO)-$Pr_{0.6}Sr_{0.4}Fe_{0.8}Al_{0.2}O_{3-\delta}$ (PSFAO) composite materials by using partial Al instead of Fe at PSFO. The influences of the CPO/PSFAO ratio and membrane thickness on the oxygen permeability of the newly developed membranes CPO-PSFAO are evaluated. Furthermore, Oxygen permeability and phase stability as well as $CO_2$ stability are studied systematically.

## 2. Experimental

**Synthesis of powders and membranes**

The polycrystalline samples of $x$CPO-(100-$x$)PSFAO ($x$ = 50, 60 and 75, denoted as 50CPO-50PSFAO, 60CPO-40PSFAO and 75CPO-25PSFAO, respectively) composites were synthesized via a modified Pechini method. The appropriate stoichiometric metal nitrates were sequentially dissolved in deionized water and citric acid as a chelating agent and ethylene glycol as a dispersing agent were added to the solution. The solution was heated and stirred using a magnetic stirrer to evaporate water to obtain a gel. The powder precursor was obtained after drying the gel in a furnace at 150 °C. The precursor was calcined to 600 °C to remove the organic components therein and further heated at 950 °C for 10 h to get a powder of $x$CPO-(100-$x$)PSFAO ($x$ = 50, 60 and 75). The as-prepared powder after grinding and screening was pressed to ~ 10 MPa to obtain green disks. The green disks were sintered at 1450 °C in air for 5 h to densify the $x$CPO-(100-$x$)PSFAO ($x$ = 50, 60 and 75) membranes. The desired thickness membranes can be made by polishing with sand papers. Finally, the oxygen permeation membranes were prepared for test after washing with ethanol.

**Characterization of materials**

Powder X-ray diffraction (XRD, D-MAX 2200 VPC, Rigaku with Cu Kα) was conducted to check the phase structures of the $x$CPO-(100-$x$)PSFAO ($x$ = 50, 60 and 75) samples．*In situ* XRD was carried out in an atmosphere of air and 50 vol.% $CO_2$/50 vol.% $N_2$ from 30 °C to 900 °C with a heating rate of 12 °C min$^{-1}$. Rietveld fits were fitted on the powder diffraction data by using the FULLPROF diffraction suite software [49]. The crystal structures were generated using the Vesta program [50]. The microstructures of the as-sintered samples were investigated by scanning electron microscopy (SEM, Quanta 400F, Oxford), backscattered electron microscopy (BSEM) and energy dispersive X-ray spectroscopy (EDXS). And high resolution transmission electron microscopy (HRTEM, FEI Tecnai G2 F20 operated on 200 kV) was used to study the phase structure of 60CPO-40PSFAO powder obtained after heating at 950 °C in air for 10 h.

**Oxygen permeation flux measurement**

Oxygen permeability was measured based on our previously reported method with a home-made device [51, 52]. The $x$CPO-(100-$x$)PSFAO ($x$ = 50, 60 and 75) OTMs were sealed on an alumina tube with high temperature setting glue (Huitian, China), baked at 140 °C for 10 h, and the lateral direction of the oxygen permeable membrane was also covered with high temperature setting glue to avoid the transmission of radial oxygen affecting the final measured value. The effective working area of the oxygen permeable membrane is about 0.7088 cm$^2$. Air as a feed gas, a mixture of He or $CO_2$ (49 mL min$^{-1}$) and Ne (1 mL min$^{-1}$) of internal standard gas as sweeping gas. Flow rates of all inlet gas are controlled by the mass flow meters (Sevenstar, Beijing, China) and are periodically calibrated using a soap membrane flow meter. The gaseous mixtures were analyzed using a gas chromatograph (GC, zhonghuida-A60, Dalian, China). The oxygen permeation rate can be calculated from the eq. (1).

$$J_{O2}(mL \bullet cm^{-2} \bullet min^{-1}) = (C_{O2} - \frac{C_{N2}}{4.02}) \times \frac{F}{S} \qquad (1)$$

Where $C_{O2}$ and $C_{N2}$ represent the oxygen concentration and the nitrogen concentration, respectively, and 4.02 is the ratio of the leaked nitrogen according to the theory of Kundsen diffusion. $F$ is the total flow rate of the exhaust gas calculated from the Ne concentration, and $S$ represents the effective oxygen permeability area of the $x$CPO-(100-$x$)PSFAO ($x$ = 50, 60 and 75) dual phase membranes sealed on the corundum tube [53-55]. In this study, the leakage flux of oxygen was generally less than 5% of the total oxygen flux.

**Result and Discussion**

The XRD patterns of $x$CPO-(100-$x$)PSFAO ($x$ = 50, 60 and 75) composite powders after calcined at 950 °C for 10 h are shown in **Fig. 1**. All three composite powders consist of only CPO and PSFAO phases, suggesting that the $x$CPO-(100-$x$)PSFAO ($x$ = 50, 60 and 75) samples can be successfully synthesized via the modified Pechini method. In order to check phase structures of the $x$CPO-(100-$x$)PSFAO ($x$ = 50, 60 and 75) composite membranes, the sintered membranes are also characterized by XRD (see **Fig. S1)**. The results reveal that all three sintered composite membranes under studied also consist of only CPO and PSFAO phases, which further confirm that $x$CPO-(100-$x$)PSFAO dual phase membranes have been prepared successfully.

Further, the unit cell parameters of CPO and PSFAO in three composites are compared. With Rietveld refinement, the unit cell parameters of the CPO and PSFAO two phases in $x$CPO-(100-$x$)PSFAO ($x$ = 50, 60 and 75) composites are obtained and summarized in **Table 1**. The Rietveld analysis results reveal that the pure phase CPO exhibits a cubic fluorite structure with space group of $Fm\overline{3}m$ (No. 225, $a = b = c = 5.4050(3)$Å), while the pure phase PSFAO exhibits an orthorhombic distorted perovskite structure with space group of $Imma$ (No. 74, $a = 5.4412(3)$Å, $b = 7.7150(6)$Å, $c = 5.4770(3)$Å) at room temperature. There is no big difference between the unit cell parameters of the CPO and PSFAO two phases in three $x$CPO-(100-$x$)PSFAO ($x$ = 50, 60 and 75) composites, they are all very close to those of pure CPO and PSFAO. For comparison, the pure phase PSFO and PSFAO were also synthesized by a modified Pechini method, as shown in **Fig. S2**. The lattice constants of PSFAO (No. 74, $a$ = 5.4412(3)Å, $b$ = 7.7150(6)Å, $c$ = 5.4770(3)Å) were smaller than those of pristine PSFO (No. 74, $a$ = 5.4828(2)Å, $b$ = 7.7855(8)Å, $c$ = 5.4855(4)Å).

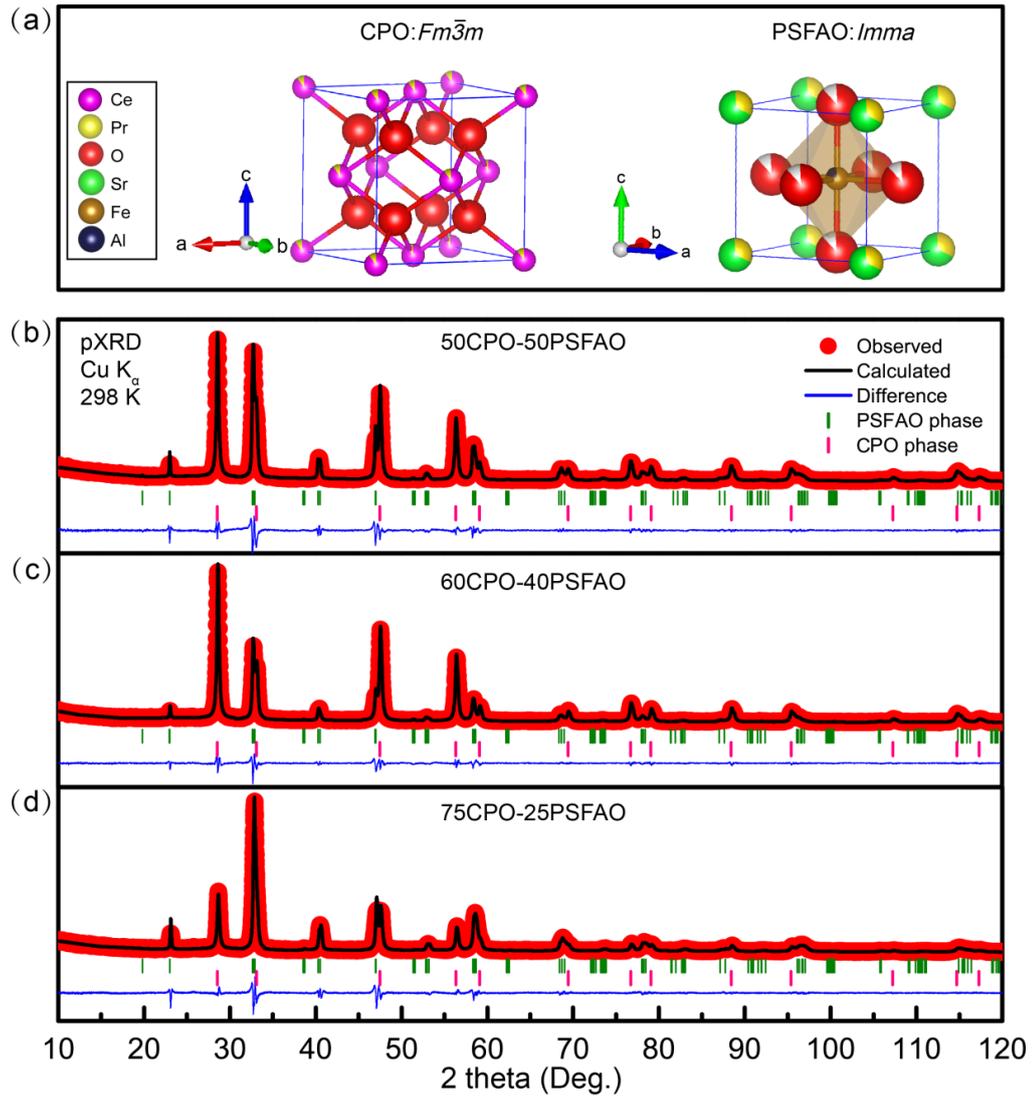

**Fig. 1.** Crystal structure characterization of CPO-PSFAO: (a) the CPO phase crystal structure on the left, the PSFAO phase crystal structure on the right; (b, c, d) Rietveld refinement XRD patterns of $x$CPO-$(100-x)$PSFAO ($x$ = 50, 60 and 75) powders after calcined at 950 °C for 10 h in air, respectively.

**Tab 1** Unit cell parameters for pure CPO, PSFAO and two phases in $x$CPO-$(100-x)$PSFAO ($x$ = 50, 60 and 75) composites.

| Phase CPO:PSFAO (wt.%) | Powder | | | | Membrane | | | |
|---|---|---|---|---|---|---|---|---|
| | CPO | PSFAO | | | CPO | PSFAO | | |
| | a = b = c(Å) | a(Å) | b(Å) | c(Å) | a = b = c(Å) | a(Å) | b(Å) | c(Å) |
| 0:100 | - | 5.4412 | 7.7150 | 5.4770 | - | - | - | - |
| 50:50 | 5.4106 | 5.4392 | 7.7317 | 5.4831 | 5.4116 | 5.4357 | 7.7333 | 5.4846 |
| 60:40 | 5.4101 | 5.4414 | 7.7356 | 5.4844 | 5.4110 | 5.4421 | 7.7348 | 5.4860 |
| 75:25 | 5.4118 | 5.4377 | 7.7279 | 5.4816 | 5.4118 | 5.4384 | 7.7304 | 5.4865 |
| 100:0 | 5.4050 | - | - | - | - | - | - | - |

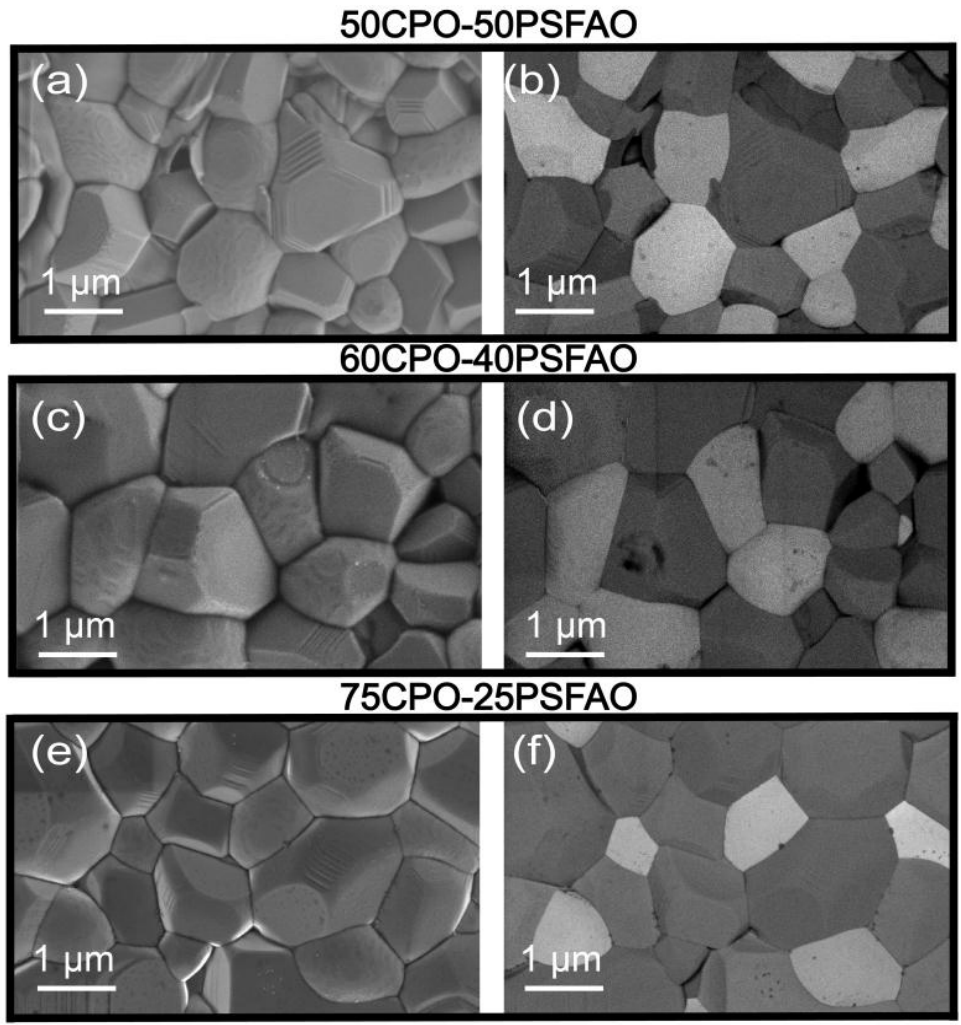

**Fig. 2.** SEM (left line, a, c, e), and BSEM (right line, b, d, f) micrographs of the surfaces of 50CPO-50PSFAO (a, b), 60CPO-40PSFAO (c, d) and 75CPO-25PSFAO (e, f) dual-phase membranes after sintering at 1450 °C for 5 h. In BSEM, the dark grains represent the PSFAO grains, the light ones represent the CPO grains, since the contribution of the backscattered electrons to the SEM signal intensity is proportional to the atomic number.

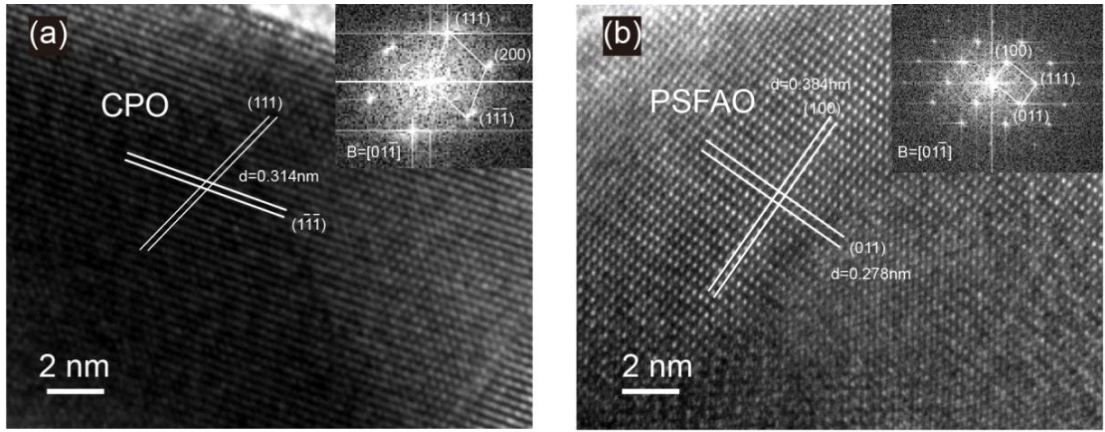

**Fig. 3.** HRTEM images of CPO and PSFAO in the dual-phase 60CPO-40PSFAO powder, the characteristic (111), (1$\bar{1}\bar{1}$) CPO and (100), (011) PSFAO in the 60CPO-40PSFAO powder. Inset the upper right corner of **Fig 3a, b** shows the [0$\bar{1}\bar{1}$] zone axis pattern (ZAP) of CPO and the [0$\bar{1}\bar{1}$] ZAP of PSFAO, respectively.

Next, we inspected the microstructure of the $x$CPO-(100-$x$)PSFAO ($x$ = 50, 60 and 75) membranes by SEM, BSEM and HRTEM. **Fig. 2** shows the SEM and BSEM of the surfaces in the sintered $x$CPO-(100-$x$)PSFAO ($x$ = 50, 60 and 75) composites. No cracks or pin holes are observed in the SEM images in **Fig. 2(a, c, e)**, which indicate that all three sintered membranes are dense. The distribution of CPO and PSFAO two phases is highly dependent on the CPO/PSFAO ratio, but the particle sizes isn't. The average grain size of PSFAO is close to that of CPO. Further, the CPO and PSFAO grains could be identified by BSEM **Fig. 2(b, d, f)** and EDXS (**Fig. S3**), which reveal that the dark grains in BSEM (green in EDXS) are PSFAO and the light ones in BSEM belong to CPO (yellow color in EDXS). **Fig. S3** presents the corresponding element mapping images. Two phases distributed uniformly and no agglomeration of one phase was observed, which signify that CPO and PSFAO two phases has a good compatibility during sintering. More detailed information on the crystal structures of CPO and PSFAO was obtained from HRTEM characterization. **Fig. 3a, b** show the HRTEM images of CPO and PSFAO in the 60CPO-40PSFAO composite, the characteristic (111), (1$\bar{1}\bar{1}$) CPO and (100), (011) PSFAO in the 60CPO-40PSFAO powder. Inset the upper right corner of **Fig. 3a, b** shows the [0$\bar{1}\bar{1}$] zone axis pattern (ZAP) of CPO and the [0$\bar{1}\bar{1}$] ZAP of PSFAO, respectively. It indicates CPO phase has cubic structure;

the PSFAO has orthorhombic structure, which is consistent in XRD data previously mentioned.

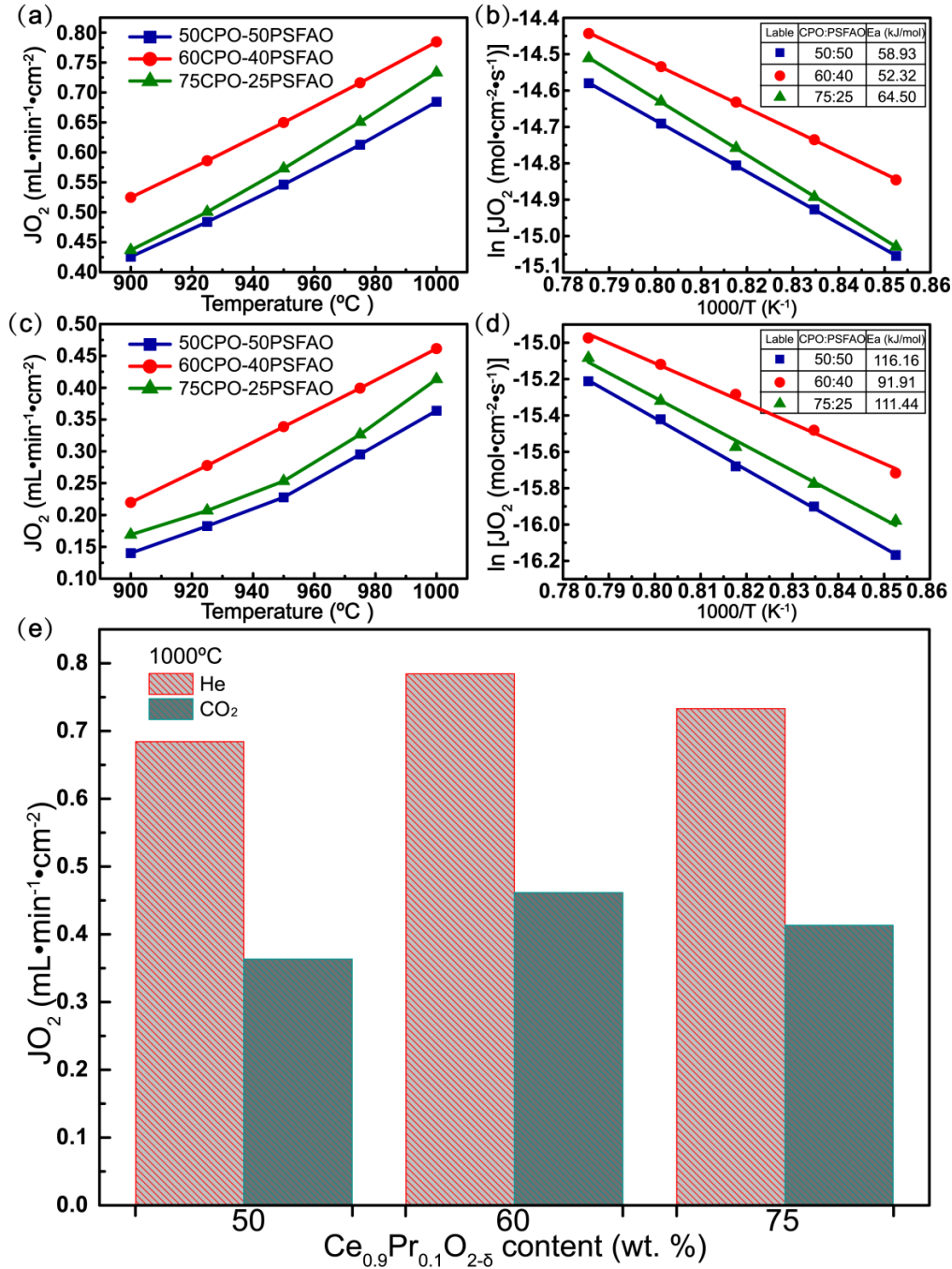

**Fig. 4.** Oxygen permeation fluxes (a, c) and its Arrhenius plot (b, d) through the $x$CPO-(100-$x$)PSFAO ($x$ = 50, 60, 75) dual-phase membranes as a function of temperature with He (a, b) or $CO_2$ (c, d) as sweep gas and (c) Oxygen permeation flux of $x$CPO-(100-$x$)PSFAO ($x$ = 50, 60, 75) dual-phase membranes after sintering at 1450 °C for 5 h as a function of CPO content at 1000 °C with He or $CO_2$ as sweep gas. *Condition: 150 mL min$^{-1}$ air as the feed gas, 49 mL min$^{-1}$ He or 49 mL min$^{-1}$ $CO_2$ as sweep gas, 1 mL min$^{-1}$ Ne as an internal standard gas. Membrane thickness: 0.6 mm.*

To study the effect of CPO/PSFAO ratio on the oxygen permeability, three different CPO-PSFAO composites have been prepared. **Fig. 4a** exhibits the oxygen permeation fluxes ($JO_2$) through the 0.6 mm- thickness $x$CPO-(100-$x$)PSFAO ($x$ = 50, 60 and 75) membranes as a function of temperature at air/He or air/$CO_2$ condition. The oxygen permeation fluxes through all the $x$CPO-(100-$x$)PSFAO ($x$ = 50, 60 and 75) composites increase with increased temperature. In addition, it can be clearly observed that 60CPO-40PSFAO possess the highest oxygen permeability from 900 °C to 1000 °C among the studied composites. It can be considered that when the ratio of CPO:PSFAO is 60:40, the oxygen ion channel and the electron channel of the omposite material are more unblocked, compared with the case of 50:50 and 75:25. In other words, 60CPO-40PSFAO has better ionic conductivity and electronic conductivity than 50CPO-50PSFAO and 75CPO-25PSFAO. To further understand the electronic percolation for dual-phase membrane, we calculated the volume ratios of CPO-PSFAO as follows: 50wt.%-50wt.%, 60wt.%-40wt.% and 75wt.%-25wt.% correspond to 43.87vol.%-56.13vol.%, 53.97vol.%-46.03vol.% and 70.10vol.%-29.90vol.%, respectively. Videlicet, when the volume ratio of CPO to PSFAO is 53.87 : 46.03, the dual-phase membrane has the best oxygen permeability. It is very important to choose the proper ratio of fluorite to perovskite to obtain excellent oxygen ion conductivity and electronic conductivity of the composites. In the case of $x$CPO-(100-$x$)PSFAO composites, 60CPO-40PSFAO is undoubtedly the best choice. At 1000 °C, the oxygen permeation rate of 0.6 mm-thickness 60CPO-40PSFAO membrane can reach as high as 0.78 mL cm$^{-2}$ min$^{-1}$ and 0.46 mL cm$^{-2}$ min$^{-1}$ in air/He and air/$CO_2$ gradients, respectively. The activation energies can be extracted from the date in **Fig. 4b, d**. It is worth mentioning that a single activation energy was obtained no matter using He or $CO_2$ as sweep gas in the range of 900-1000 °C. Based on the previous study [56], a single activation energy is a key characteristics for a good operational stability for MIEC membranes. 60CPO-40PSFAO showed the lowest activation energy (91.91 kJ/mol), which means the lowest energy required for oxygen ions to break through the potential barrier for diffusion, resulting in the optimal oxygen permeability performance of the membranes. In addition, according to the Wagner equation[57, 58], we calculated the ionic conductivity of the $x$CPO-(100-$x$)PSFAO ($x$ = 50, 60 and 75) through oxygen permeability, which was 0.0154, 0.0179 and 0.0163 S·cm$^{-1}$ respectively. The *in situ* XRD results shown in **Fig. 6 and Fig. 11** also have verified that the 60CPO-40PSFAO sample exhibits good phase stability under our measuring temperature range. The optimal 60CPO-40PSFAO composite show the smallest of activation energy in both He and $CO_2$ atmospheres. In addition, the values of activation energy for all three composites at air/He condition are much smaller than those at air/$CO_2$ condition, signifying that stronger absorption of $CO_2$ gas than that of helium gas occur on the surfaces of the membranes. It can explain the phenomenon on the drop of the oxygen permeation rates if using $CO_2$ instead of He as sweep gas.

The optimal 60CPO-40PSFAO composition was chosen for further study of the thickness effect on the oxygen permeation rates through these 60CPO-40PSFAO composites using helium as sweep gas. As shown in **Fig. 5a**, it is obvious that the thinner the 60CPO-40PSFAO sample, the higher oxygen permeation rate of the 60CPO-40PSFAO is obtained at the same temperature. For instance, at 1000 °C, the order of $JO_2$ through the 60CPO-40PSFAO membrane of different thicknesses at air/He condition from high to low is as follows: 0.33 mm (1.03 mL cm$^{-2}$ min$^{-1}$) > 0.4 mm (0.95 mL cm$^{-2}$ min$^{-1}$) > 0.6 mm (0.78 mL cm$^{-2}$ min$^{-1}$) > 0.8 mm (0.65 mL cm$^{-2}$ min$^{-1}$). **Fig. 5b** presents the Arrhenius plots of the oxygen permeation fluxes through 60CPO-40PSFAO membranes with various thicknesses under an air/He atmosphere. According to **Fig. 5b**, the apparent activation energies of 0.33 - 0.8 mm-thickness 60CPO-40PSFAO in the range of 900-1000 °C were calculated to be 68.60, 67.97, 52.34, and 70.10 kJ mol$^{-1}$, respectively.

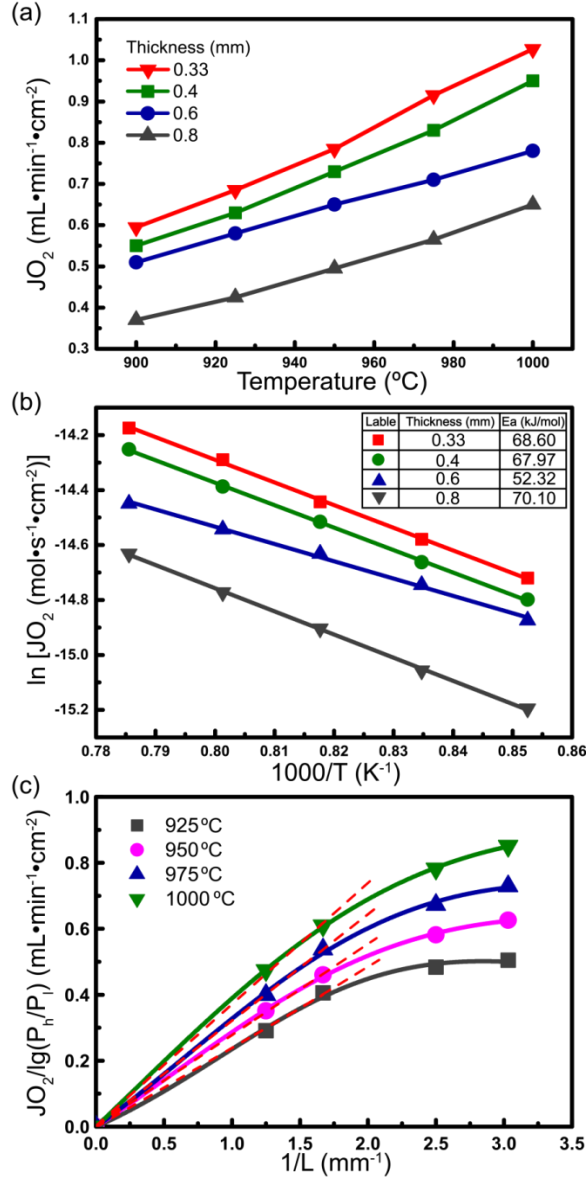

**Fig. 5.** Oxygen permeation fluxes (a) and its Arrhenius plot (b) through the 60CPO-40PSFAO dual-phase membranes with different thickness; and (c) Oxygen permeation fluxes as a function of reciprocal thickness for different temperatures with pure He as sweep gas. *Condition: 150 mL min$^{-1}$ air as the feed gas, 49 mL min$^{-1}$ He, 1 mL min$^{-1}$ Ne as an internal standard gas. Membrane thickness: 0.33 mm, 0.4 mm, 0.6 mm and 0.8 mm.*

For better understanding the critical thickness, the oxygen permeation flux $J_{O_2}/lg(P_h/P_l)$ versus $1/L$ at 925, 950, 975 and 1000 °C, is plotted in **Fig. 5c** (data extracted from **Fig. 5a**). It is clear that when the thickness is larger than 0.6 mm, $J_{O_2}/lg(P_h/P_l)$ belongs to a linear trend; on the contrary, the linear relationship disappears. This can be explained by the eq. (2) [58, 59]:

$$J_{O2}(\text{mL cm}^{-2}\text{ min}^{-1}) = \frac{\langle C_h D_h \rangle}{4L} \ln \frac{P_h}{P_l} \qquad (2)$$

where $J_{O2}$ is the oxygen permeation rate, $C_h$ and $D_h$ are the concentration of oxygen vacancies and the diffusion coefficient of oxygen vacancies. $P_h$, $P_l$, $L$ present the high oxygen partial pressure on the feed side, the low oxygen partial pressure on the sweep side, and thickness of membrane, respectively.

According to the eq. (2), if the oxygen transport through a dense MIEC membrane is mainly controlled by the oxygen bulk diffusion, the $JO_2$ versus $L$ obeys the inverse proportional relation at an invariable oxygen partial pressure gradient. Contrarily, the surface oxygen exchange would become a key factor when the membrane thickness reduces to small enough. From eq. (2), we can expect that a plot of $JO_2/lg(P_h/P_l)$ against $1/L$ gives a straight line. However, according to **Fig. 5c**, the oxygen permeation flux deviates from the linear trend and increases very slowly when the thickness is below 0.6 mm. It can be concluded that the surface exchange process is not negligible when the membrane thickness is below 0.6 mm, while the oxygen bulk diffusion mainly controls the overall oxygen transport process when the thickness is greater than 0.6 mm. And when the membrane thickness is less than 0.3 mm, the oxygen permeation process is mainly controlled by interfacial exchange. $L_c$ is defined as the point at which the transition occurs from predominantly diffusion-controlled to control by interfacial exchange[60], so $L_c$ of 60CPO-40PSFAO membrane is between 0.3mm and 0.6mm.

Further examination on the phase stability of CPO-PSFAO at high temperatures was achieved from the *in situ* XRD measurements (**Fig. 6a and b**). No signature of impure peaks was observed no matter in air or in $CO_2$ containing atmospheres, which indicate that the 60CPO-40PSFAO exhibits a good structure stability in both air and $CO_2$ atmospheres. We next consider the phase stability of the 60CPO-40PSFAO in low oxygen partial pressure. **Fig. 7** presents the XRD patterns of 60CPO-40PSFAO powders after heating at 950 °C in pure Ar for different times. No additional peaks except for CPO and PSFAO two phases can be detected by XRD, signifying the good structural stability under Ar. The above results suggest that 60CPO-40PSFAO membrane has good thermal and chemical stability in air or low oxygen partial pressure atmosphere.

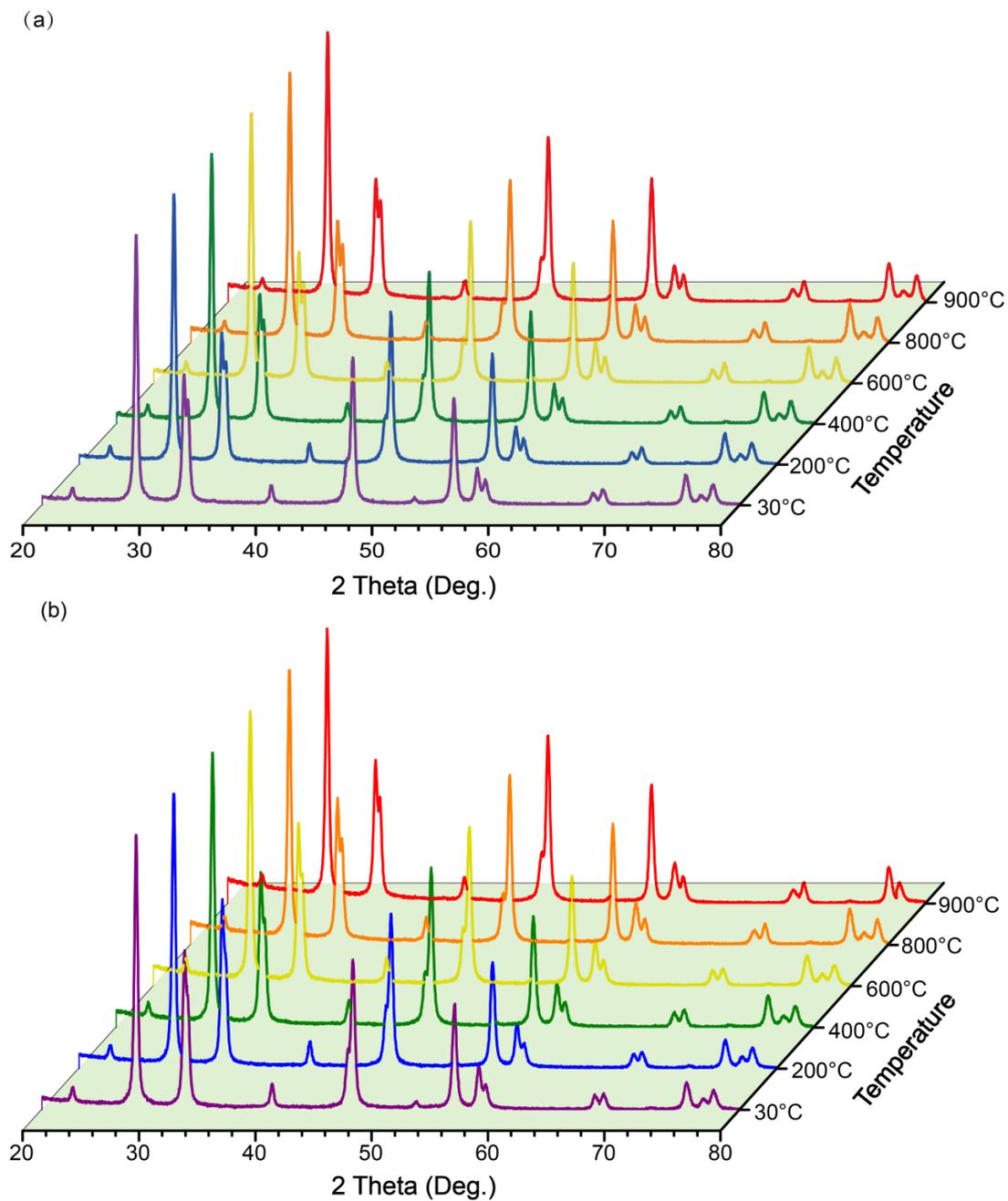

**Fig. 6.** *In situ* XRD patterns of 60CPO-40PSFAO powder calcined at 950 °C for 10 h in air for increasing temperature (a) under air atmosphere and (b) under 50 vol.% $CO_2$ + 50 vol.% $N_2$, heating rate 10 °C/min, holding 30 min before each temperature test.

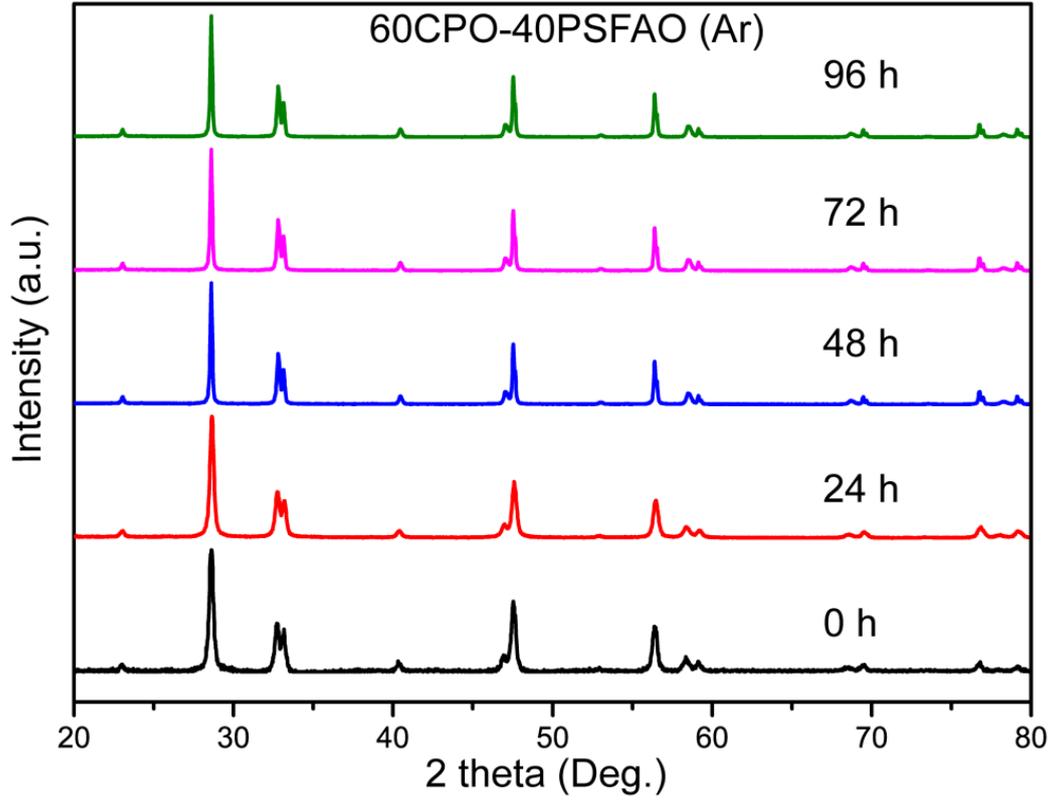

**Fig. 7.** XRD patterns of 60CPO-40PSFAO dual-phase membrane powders after exposed to pure Ar for different hours.

In addition, the unit cell parameters obtained from the *in situ* XRD data in air and $CO_2$-containing atmospheres are plotted in **Fig. 8.** All the unit cell parameters enlarge with enhancing temperatures due to the thermal expansion. Further, we can obtained the thermal expansion coefficients (TEC) by following the definition $\frac{dV}{dT} \times \frac{1}{V_0}$ (V is the unit cell volume and $V_0$ is the unit cell volume at room temperature) [59]. It was found that the obtained TEC of CPO is $5.2 \times 10^{-5}$ $K^{-1}$ and close to that of PSFAO ($4.5 \times 10^{-5}$ $K^{-1}$), which should assure the mechanical stability of the CPO-PSFAO composite during thermal cycling.

We also examined the reversibility of the oxygen permeability of the 60CPO-40PSFAO composite. **Fig. 9** illustrates the changeability of the oxygen permeability of the 0.6 mm-thickness 60CPO-40PSFAO sample by periodically adjusting $CO_2$ and He sweep gases at 1000 °C. A high oxygen permeation rate of 0.79 mL $cm^{-2}$ $min^{-1}$ can be achieved when adjusting He as sweep gas; while the oxygen permeation rate decline to

0.46 mL cm$^{-2}$ min$^{-1}$ when adjusting CO$_2$ as sweep gas.

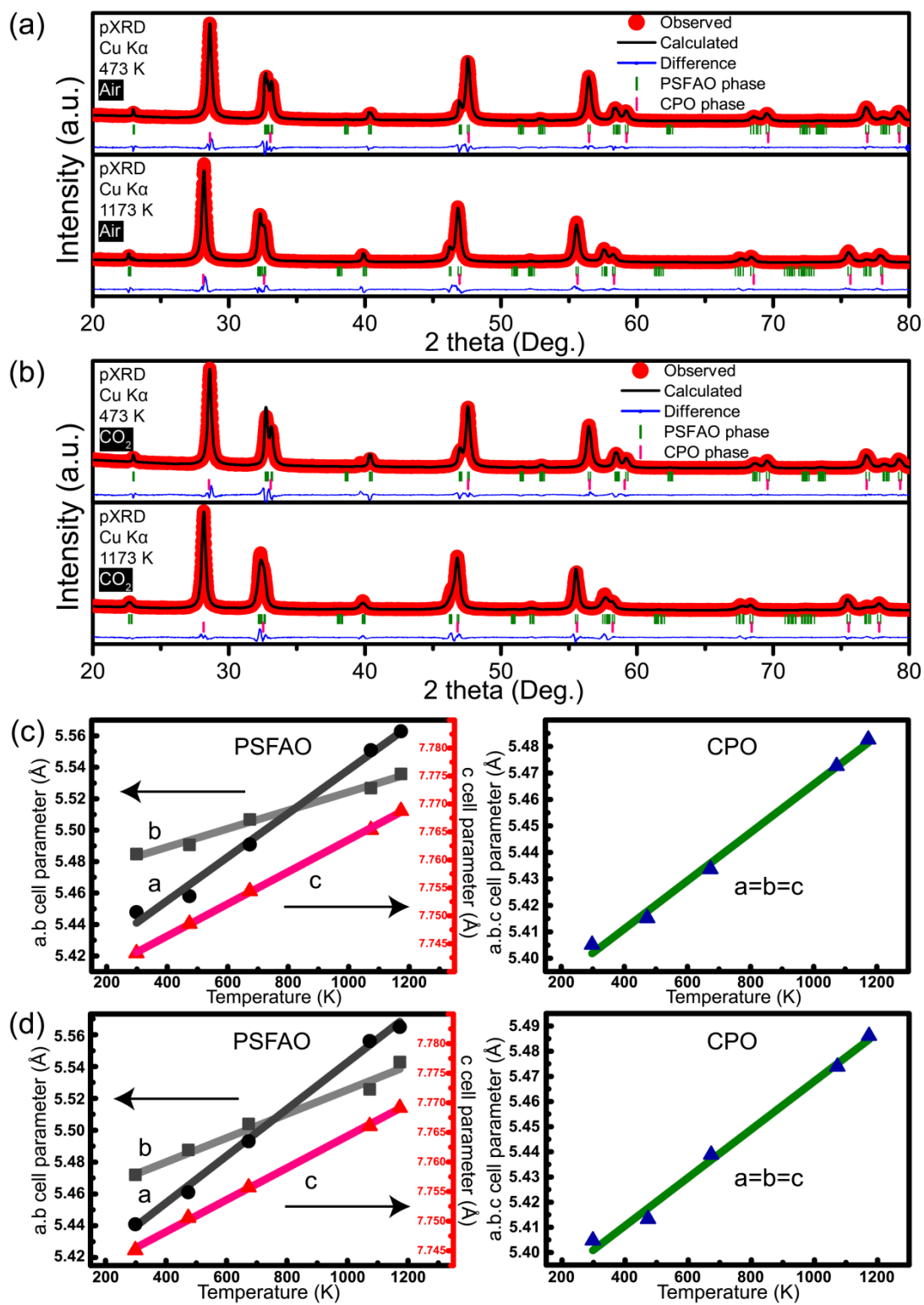

**Fig. 8.** *In situ* XRD patterns of 60CPO-40PSFAO powder (a) in air, and (b) in 50 vol.% CO$_2$+50 vol.% N$_2$. Unit cell parameter as a function of temperature for (c) in air and (d) in 50 vol.% CO$_2$ +50 vol.% N$_2$ atmosphere.

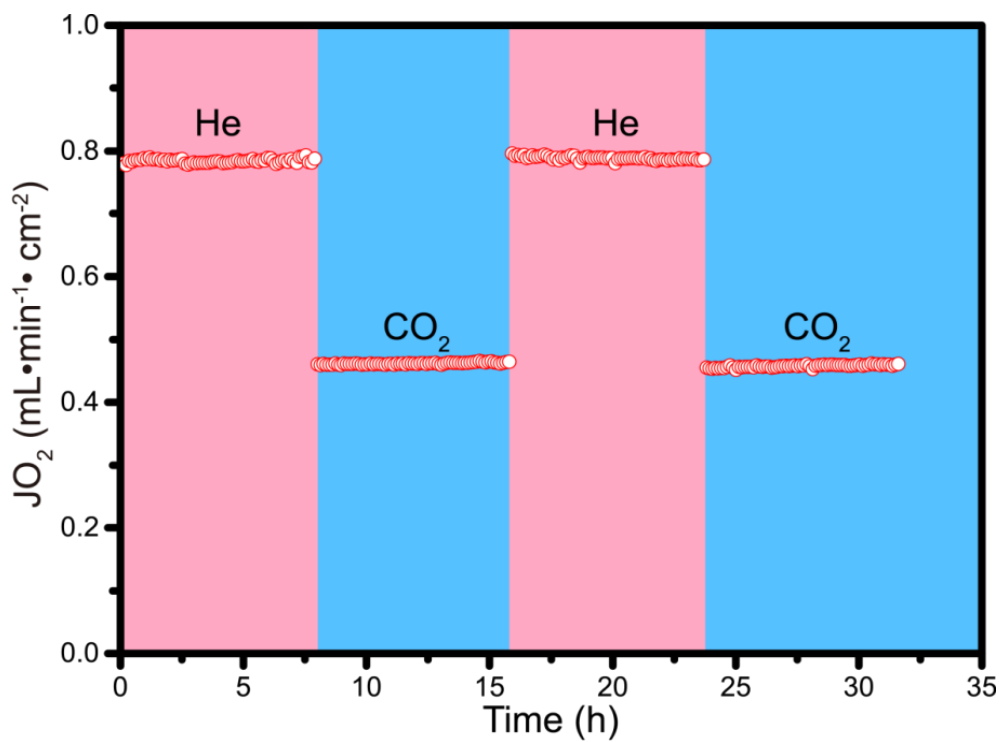

**Fig. 9.** Oxygen permeation flux of 60CPO-40PSFAO membrane while periodically changing the sweep gas between He and $CO_2$ at 1000 °C. *Condition: 150 mL min$^{-1}$ air as the feed gas, 49 mL min$^{-1}$ He or 49 mL min$^{-1}$ $CO_2$ as sweep gas, 1 mL min$^{-1}$ Ne as an internal standard gas. Membrane thickness: 0.6 mm.*

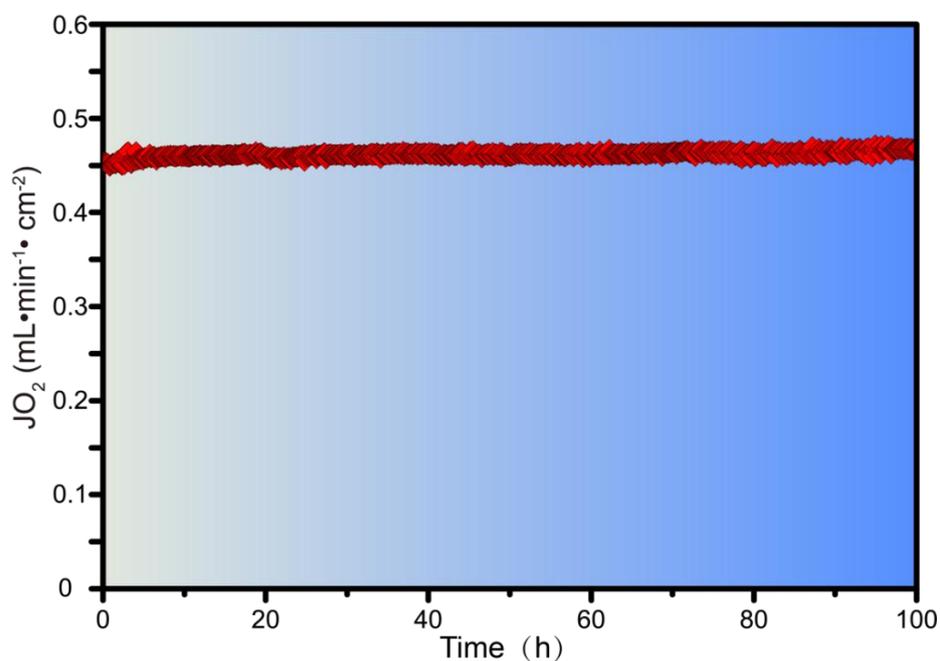

**Fig. 10.** Oxygen permeation flux of 60CPO-40PSFAO membrane as a function of time using pure $CO_2$ as sweep gas at 1000 °C.
*Condition: 150 mL min$^{-1}$ air as the feed gas, 49 mL min$^{-1}$ $CO_2$ as sweep gas, 1 mL min$^{-1}$ Ne as an internal standard gas. Membrane thickness: 0.6 mm.*

This phenomenon differs from the previously reported behavior on the single perovskite-type membranes, with no permeation immediately or a rapid decline of the oxygen flux due to the formation of carbonate [25, 28, 61, 62]. For instance, Jian Song et al. investigated the $CO_2$ erosion on the single perovskite $BaCo_{0.85}Bi_{0.05}Zr_{0.1}O_{3-\delta}$ hollow fiber membranes and found no permeation was observed when even 10 vol.% $CO_2$ in the feed or sweep sides due the formation of $BaCO_3$ carbonate on the surface [25, 28, 61]. Here, the cause of slight drop of the oxygen permeation rate due to the adsorption effect of $CO_2$ on surface of the sample but not the carbonates formation [47].

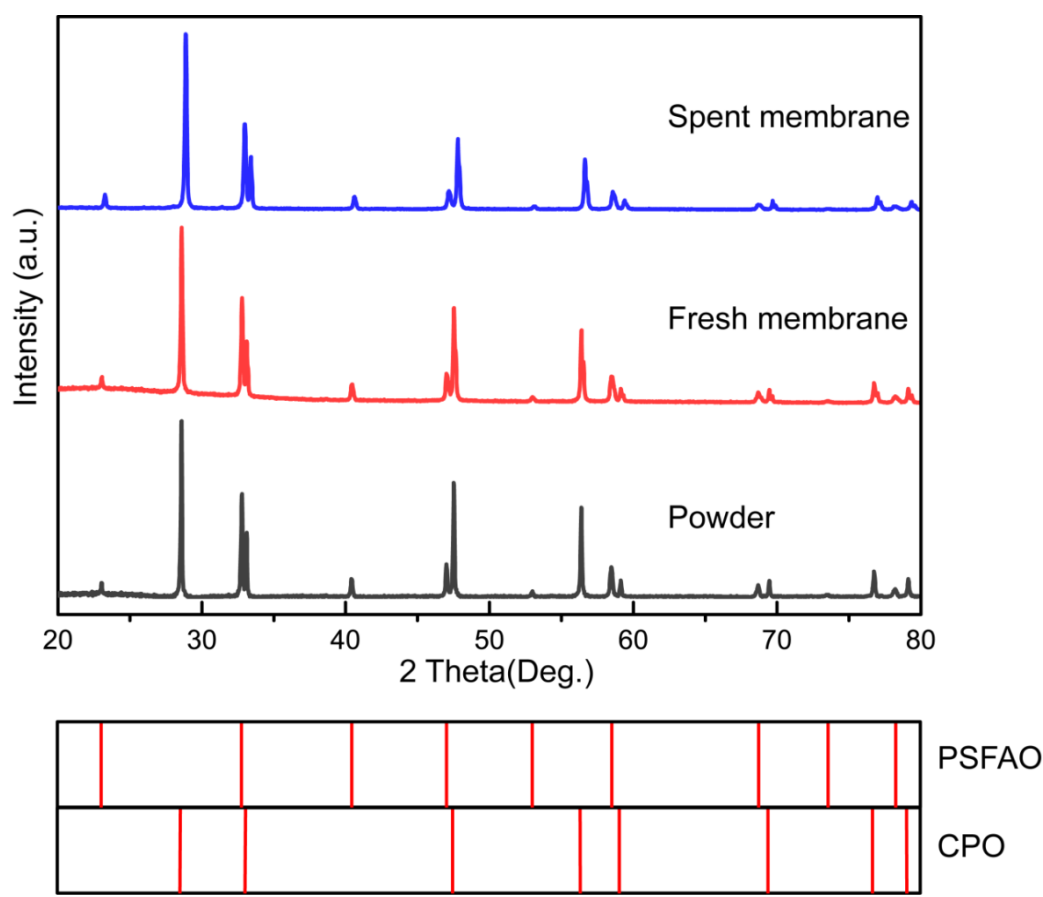

**Fig. 11.** The XRD patterns of fresh and spent 60CPO-40PSFAO dual-phase membranes after 100 h test under $CO_2$ atmosphere.

To gain more insight into the long-term operational oxygen permeability and $CO_2$ stability, we operated the 0.6 mm-thickness 60CPO-40PSFAO membrane under

air/CO$_2$ gradient for at least 100 h. The time-dependent oxygen permeation rates through the 60CPO-40PSFAO samples with pure CO$_2$ as sweep gas at 1000 °C was shown in **Fig. 10**. The oxygen permeation flux increase slightly from 0.45 to 0.46 mL cm$^{-2}$ min$^{-1}$ and no degradation was observed. In order to further check the stability of the 60CPO-40PSFAO membrane, XRD was used to check the phase structures of both sides of the spent membrane after the long-term oxygen permeation measurements with pure CO$_2$ as sweep gas. As shown in **Fig. 11**, no impurities phases were observed for the spent 60CPO-40PSFAO sample, further confirming that the 60CPO-40PSFAO composite has good operating stability under CO$_2$ atmosphere.

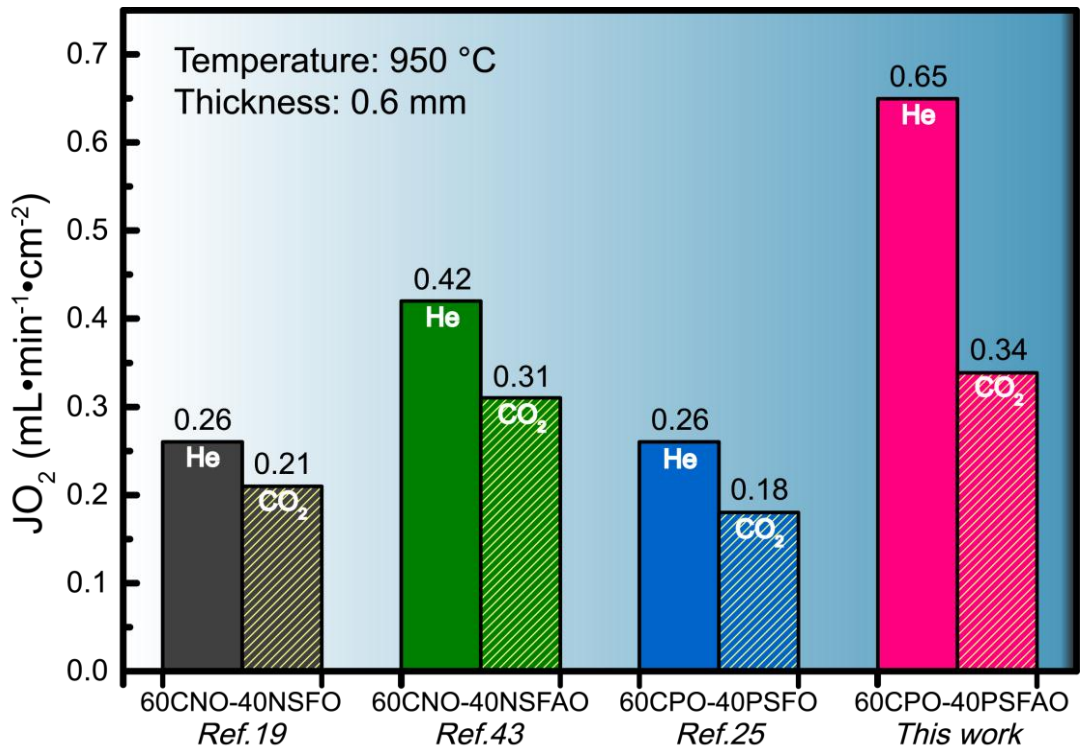

**Fig. 12.** The oxygen permeation fluxes of several types of CO$_2$-stable dual-phase membranes.

**Fig. 12** compares the oxygen permeability of the pristine and Al-doping CO$_2$-stable composites. Obviously, the oxygen permeation fluxes through both 60CNO-40NSFO and 60CPO-40PSFO pristine membranes are smaller than those of Al-doping 60CNO-40NSFAO and 60CPO-40PSFAO membranes. This is attributed to the fact that the partial replacement of $Fe^{3+}/Fe^{4+}$ by $Al^{3+}$ increases the oxygen vacancy of the material, which makes the material have better ionic conductivity. By comparison, we can find that the 60CPO-40PSFAO shows nearly 2.5 and 1.6 times higher than those of pristine 60CPO-40PSFO sample under air/He and air/$CO_2$ gradients, respectively, in the same measuring condition. In addition, it can be seen that our 60CPO-40PSFAO membrane shows higher oxygen permeability compared with that of 60CNO-40NSFAO. These results indicate that our 60CPO-40PSFAO composite with excellent oxygen permeability as well as good $CO_2$ tolerance, which is likely to be used for $CO_2$ capture based on oxy-fuel concept.

## 4. Conclusions

Composites based on $x$Ce$_{0.9}$Pr$_{0.1}$O$_{2-\delta}$(CPO)-(100-$x$)Pr$_{0.6}$Sr$_{0.4}$Fe$_{0.8}$Al$_{0.2}$O$_{3-\delta}$(PSFAO) ($x$ = 50, 60 and 75) have been synthesized successfully via a modified Pechini method. XRD and SEM-EDXS analysis confirmed that CPO and PSFAO two phases have good compatibility in all three composites. Oxygen permeability test results reveal that the CPO/PSFAO ratio remarkably affects the oxygen permeation flux (60CPO-40PSFAO > 75CPO-25PSFAO > 50CPO-50PSFAO). At 1000 °C, the 0.33mm-thickness 60CPO-40PSFAO composite yields a high oxygen flux of 1.03 mL cm$^{-2}$ min$^{-1}$ at air/He gradient, which is over the general requirement value (1 mL cm$^{-2}$ min$^{-1}$) for oxygen transport membrane technology materials in industries. Moreover, the optimal 60CPO-40PSFAO sample can work stably for at least 100 h under air/$CO_2$ gradient, indicating that the 60CPO-40PSFAO composite has good $CO_2$ tolerance, recommending the 60CPO-40PSFAO material as a promising candidate for $CO_2$ capture based on the oxy-fuel concept. The oxygen permeability and stability of the samples have been both improved by chemical doping Al into the perovskite phase. Finally, this paper is likely to provide some reference for the research of other dual phase membrane systems.

**Acknowledgment**

H. X. Luo acknowledges the financial support by "Hundred Talents Program" of the Sun Yat-Sen University and National Natural Science Foundation of China (21701197). M.R. Li is supported by the he "One Thousand Youth Talents" Program and the National Natural Science Foundation of China (21875287).

# Supporting information

## High CO$_2$-tolerance oxygen permeation dual-phase membranes Ce$_{0.9}$Pr$_{0.1}$O$_{2-\delta}$-Pr$_{0.6}$Sr$_{0.4}$Fe$_{0.8}$Al$_{0.2}$O$_{3-\delta}$

Lei Shi[a], Shu Wang[a], Tianni Lu[b], Yuan He[a], Dong Yan[a], Qi Lan[a], Zhiang Xie[a], Haoqi Wang[a], Man-Rong Li[c], Juergen Caro[d], Huixia Luo[a]*

[a]School of Material Science and Engineering and Key Lab Polymer Composite & Functional Materials, Sun Yat-Sen University, No. 135, Xingang Xi Road, Guangzhou, 510275, P. R. China

[b]School of Materials Sciences and Engineering, Shenyang Aerospace Unversity, Shenyang, 110136, P. R. China

[c]School of Chemistry, Sun Yat-Sen University, No. 135, Xingang Xi Road, Guangzhou, 510275, China

[d]Institute of Physical Chemistry and Electrochemistry, Leibniz University of Hannover, Callinstr. 3A, D-30167 Hannover, Germany

*Corresponding author/authors complete details (Telephone; E-mail:)   (+0086)-2039386124

luohx7@mail.sysu.edu.cn

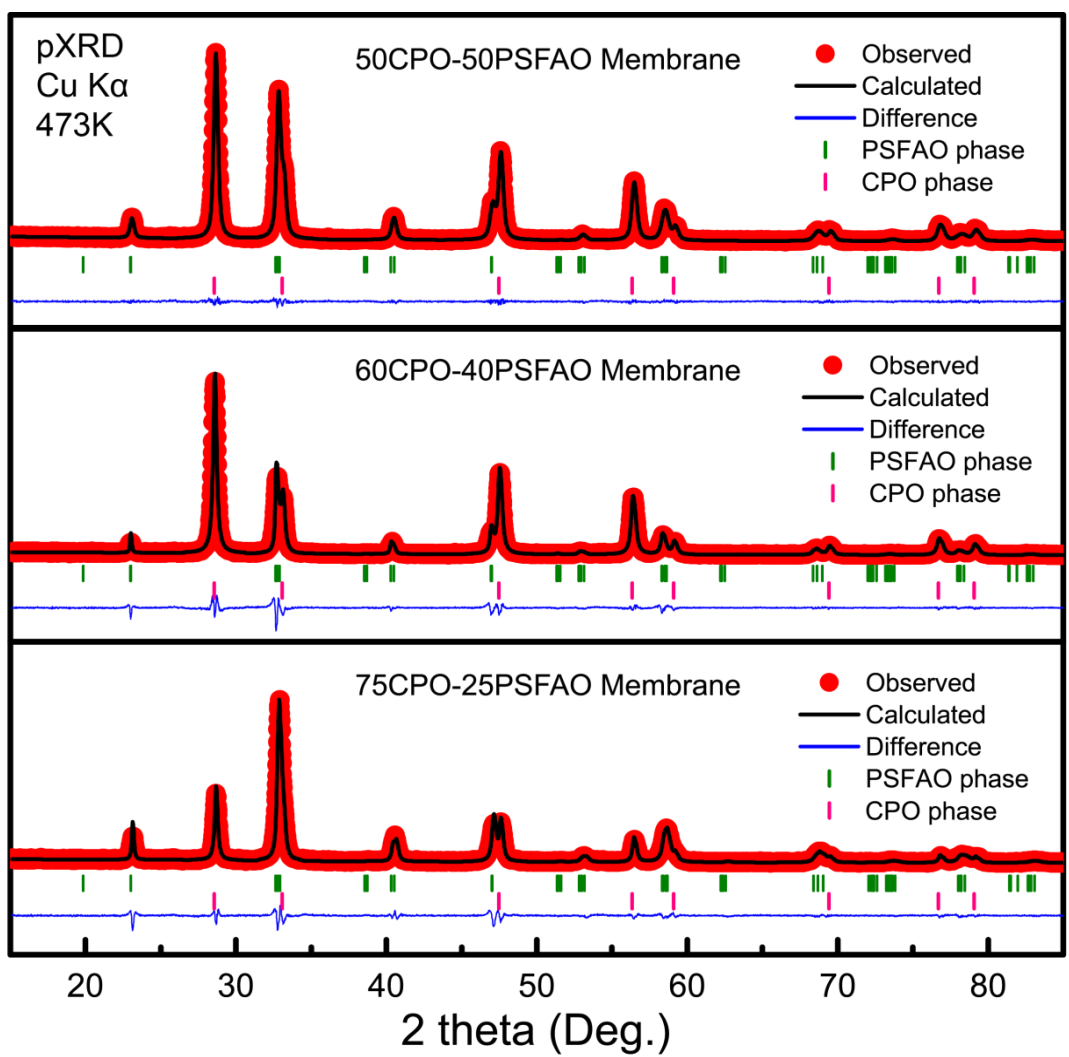

**Fig. S1.** Rietveld refinement XRD patterns of $x$CPO-(100-$x$)PSFAO ($x$ = 50, 60 and 75) dual-phase membranes after sintered at 1450 °C for 5 h in air, respectively.

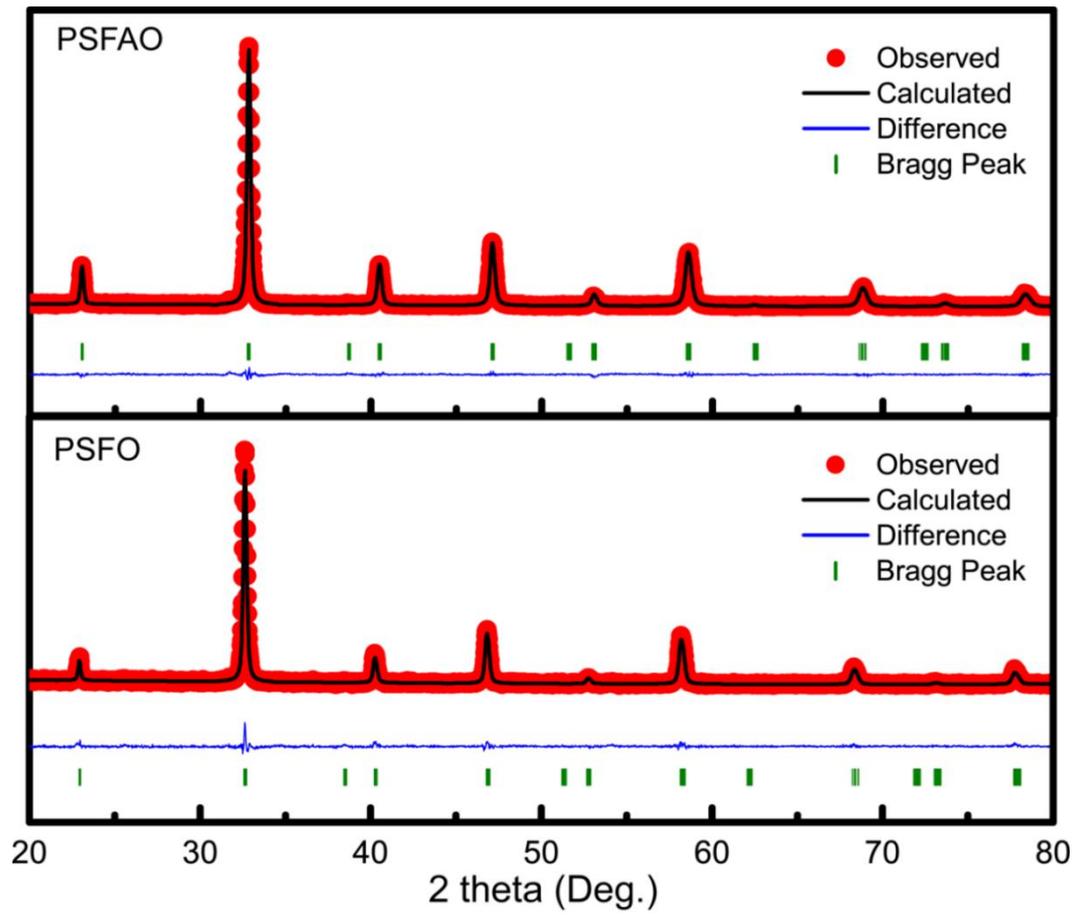

**Fig. S2.** Comparison of XRD patterns of PSFAO and PSFO powders.

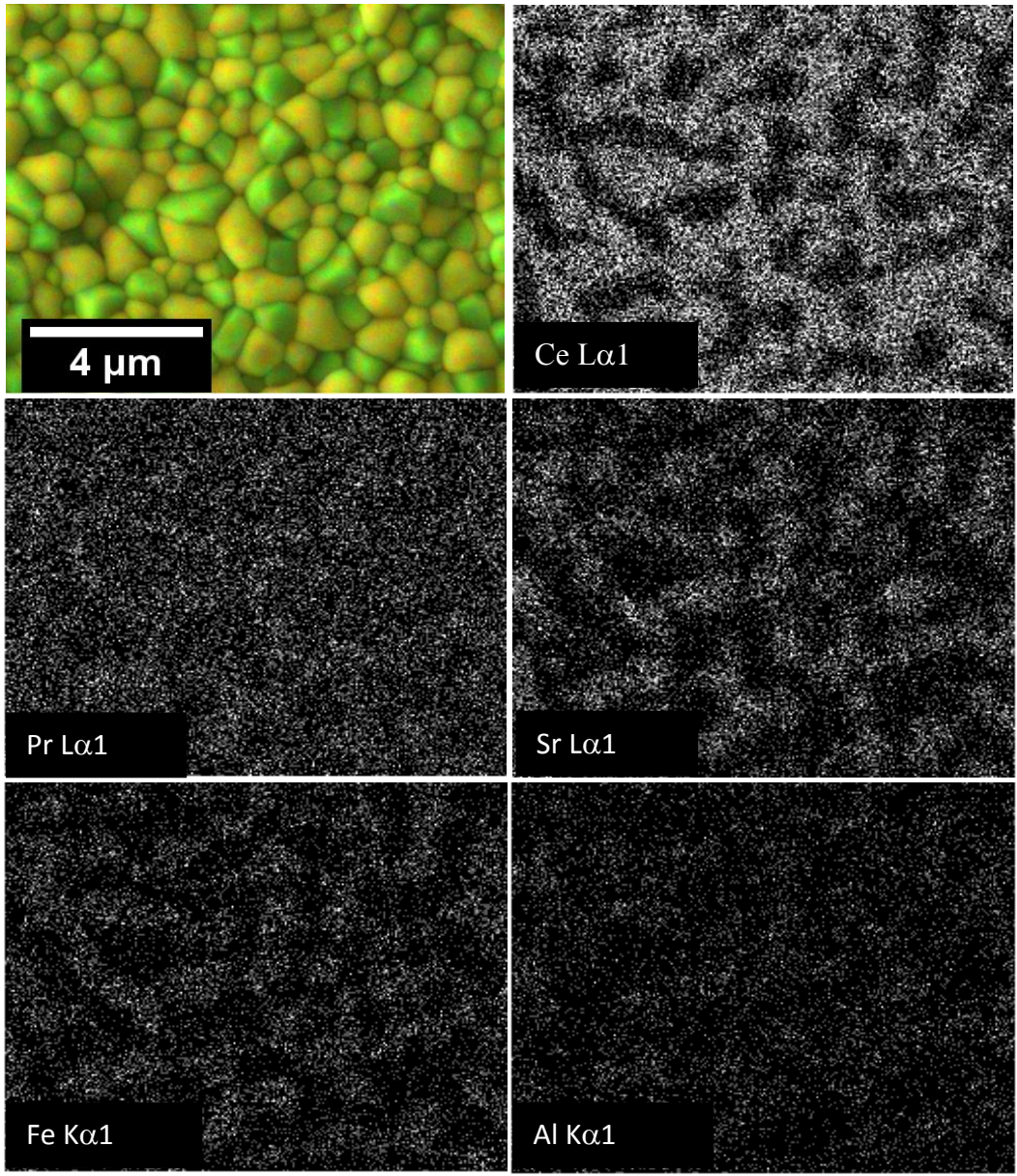

**Fig. S3.** EDXS mappings for Ce, Pr, Sr, Fe and Al elements in the 60CPO-40PSFAO dual-phase membrane sintered at 1450 °C for 5 h.